# Particle Shape Control *via* Etching of Core-Shell Nanocrystals

*Alberto Leonardi\*, Michael Engel\**

Institute for Multiscale Simulation, Friedrich-Alexander University Erlangen-Nürnberg, Nägelsbachstraße 49b, 91052 Erlangen (Germany)

**ABSTRACT**

The application of nanocrystals as heterogeneous catalysts and plasmonic nanoparticles requires fine control of their shape and chemical composition. A promising idea to achieve synergistic effects is to combine two distinct chemical and/or physical functionalities in bi-metallic core-shell nanocrystals. Although techniques for the synthesis of single-component nanocrystals with spherical or anisotropic shape are well established, new methods are sought to tailor multi-component nanocrystals. Here, we probe etching in a controlled redox environment as a synthesis technique for multi-component nanocrystals. Our Monte Carlo computer simulations demonstrate the appearance of characteristic non-equilibrium intermediate microstructures that are further thermodynamically tested and analyzed with molecular dynamics. Convex platelet, concave polyhedron, pod, cage and strutted-cage shapes are obtained at room temperature with fully coherent structure exposing crystallographic facets and chemical elements along distinct particle crystallographic directions. We observe that structural and dynamic properties are markedly modified compared to the untreated compact nanocrystal.

Manipulation of atomic bonds is an established means towards the design of chemical or physical properties of materials at the nanoscale.[1–9] The deformation of the crystalline lattice modifies the binding energy and so the chemical activity and selectivity of metal nanocrystals. The lattice distortion of the nanocrystals is ruled by structural and microstructural features such as size, shape, and defects.[3,6,9–11] Control of nanocrystal shape is crucial to tailor the functionality of individual particles as well as to target collective phenomena in self-assembled superstructures.[6,12–19] Multi-component nanocrystals with controlled shape were engineered for bioimaging, biomedical, diagnostics, energy storage, environmental remediation, and sensing applications.[20–25] In particular, the element composition modifies the sensitivity of metal nanocrystals by affecting the plasmonic properties and so the refractive index.[26] Although synthesis techniques for single-crystalline metal nanoparticles with convex shape (regular polyhedron, rod, disk, *etc.*) are advanced,[27–31] multi-component nanocrystals with concave surfaces as well as hollowed and branched regions are difficult to achieve.[20,32–37] Working towards the realization of these non-traditional nanocrystal shapes is a promising endeavor because they constitute a particularly rich design space.[38]

A number of strategies have been proposed towards non-convex nanoscale shapes and multi-component compositions. A direct method is the clustering of several nanocrystals, which creates an open microstructure,[17] but is difficult to control. Etching and/or growth of single-crystal seed nanoparticles along preferential crystallographic directions has the advantage that it preserves structural coherence across the entire microstructure.[39] In such a process, surface chemical reactivity depends on atomic coordination in the exposed facets, the concentration of auxiliary agents (capping agents, *e.g.* cetyltrimethylammonium, Bromide anions), temperature and pH of the reactor environment. However, the time evolution of the precursors has to be tuned carefully.[40] During an etching process, nanocrystals slowly but continuously evolve through a sequence of intermediate states (or short: intermediates) to complete dissolution. Once the intended intermediates are properly formed, all chemical reactions must be arrested and the auxiliary agents removed. Cleaning the crystallite surface may damage slender or weakly stable regions due to the susceptibility of partially coordinated atoms.[41] Robust and reliable methods capable of achieving a wide set of open morphology are desirable, especially if the goal is large scale production.

In this contribution, we explore the evolution of bi-metallic core-shell single-crystal nanoparticles that are oxidatively etched in a redox environment. We imagine this procedure as a strategy to control the final composition layout of the nanocrystal surface. The motivation to focus on core-shell systems is driven by their promise for synergistic functionality due to the interplay of both composing regions.[42–47] We propose to use the outer shell as a templating media to direct the transformation progress. Since the dissolution rate of the two metals comprising the nanocrystal when exposed to the oxidative solution is distinct, the core@shell character allows a highly non-uniform progression of the etching at the free surfaces.[48] Our final goal is to design nanocrystals with defined shape (particle topology, volume density, surface-to-volume ratio, coordination of the exposed facets) that have desired mechanical and chemical properties as a result of the spatial arrangement of the two element species and their lattice structure mismatch (*e.g.*, element type and unit cell parameter).[33]

**RESULTS AND DISCUSSION**

**Non-equilibrium transformation of cube@cube nanocrystals.** We first focus on Ag@Pd and Cu@Pd nanocrystals [49–51] with a simple cube@cube geometry and thin surface shell. Both compositions are such that bonds for the shell atoms (Pd, bond energy 0.326 eV) are stronger than bonds for the core atoms (Ag or Cu, bond energies 0.238 eV and 0.295 eV, respectively). The shape of the core and the surface shell together affect the crystallographic directions the core is initially exposed to the solvent, and the chemical composition determines the progression of the etching process. Our Monte Carlo (MC) simulations reveal the transformation of the convex seed crystal into concave branched hexapod microstructures with legs along the six $\langle h00 \rangle$ directions (Figure 1A,D). Furthermore, atoms belonging to the surface shell are restricted to the tips of the hexapod legs at the late stages of the etching process (Movies M_AgPd_cc and M_CuPd_cc).

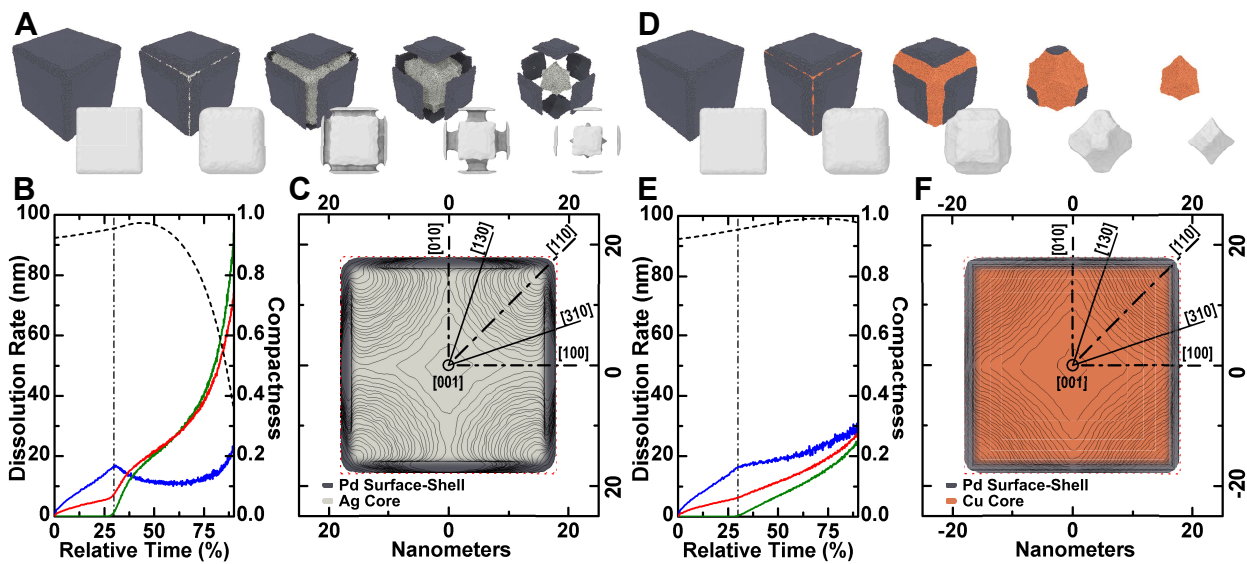

**Figure 1. Chemical etching of cube@cube nanocrystals.** Evolution of Ag@Pd (A-C) and Cu@Pd (D-F) nanoparticles exposed to the same oxidative environment (300 K, chemical potential $\mu = -2.20$ eV). A,D, Perspective and top view time-lapse snapshots from Monte Carlo (MC) simulations with element composition shown in colors (Pd dark gray, Ag light gray, Cu bronze). B,E, Compactness (right, black dashed line) and dissolution rate (left) for the whole (blue line) and the comprising crystalline regions core (red line) and surface shell (green line). The time when the core region is first exposed to the oxidative environment is indicated by a dashed vertical line. The corresponding view is shown as the second snapshot in the A,D sequences. C,F, Cross-section time-domain counter plots with characteristic crystallographic directions and element composition domains. Counter lines are spaced uniformly by 100,000 accepted MC moves.

The core and the surface shell have complementary functions in the etching progress. Differences in shape and chemical composition control the direction and the speed of the chemical reaction towards the center (Movies M_AgPd and M_CuPd). The appearance of the concave shape can be understood by analyzing the dissolution rate and compactness during the transformation process. The nanocrystal initially transforms towards a tetrahexahedron polyhedron (THH) bounded by ⟨310⟩ facets,[52,53] but then this evolution pathway changes abruptly. Once the core is exposed to the oxidative environment, its dissolution rate increases quickly, while the dissolution rate for the surface shell is proportionally depressed (Figure 1B,E). The relative dissolution rates of the core and the surface shell diverge as a function of the difference in chemical susceptibility. The larger bond energy gap of Ag@Pd (0.088 eV) compared to that of Cu@Pd (0.031 eV) leads to a stronger discrepancy in relative dissolution rate between the shell and the core. Outlines of the transforming nanocrystal collected at constant MC step intervals (100,000 accepted moves) illustrate the etching progression (Figure 1C,F).

The weaker bounded core atoms easily detach into the liquid and the etching front advances rapidly. Over time, nanocrystal compactness decreases until a concave shape emerges. While the etching of the core progresses inward, it arrests at the interface between core and surface shell. This leads to the formation of surface facets oriented towards the nanocrystal center as determined by the geometry of the core. In the last transformation stage, either surface shell platelets separate from a compact core region (Figure 1A,C) or the surface shell is fully dissolved (Figure 1D,F). Based on experimental observations,[52,53] nanocrystals typically persist for half a minute up to several minutes before they are completely dissolved. Retention times of non-equilibrium intermediates are thus expected to be at least several seconds, which is sufficient for the application of etching as a synthesis method for complex shape nanocrystals.

Temperature controls the dissolution rate of the surface atoms into the liquid solution. Because the retention time of surface atoms increases with the bond energy and the depth of shape concavities, the temperature effect is smaller for pure Pd than for Cu@Pd, which in turn is smaller than for Ag@Pd nanocrystals (Figure 2 and Movie M_Thermal). For single-component nanocrystals (pure Pd nanocubes), the shape evolution pathway is not affected by temperature. In this case, the MC acceptance probability changes with temperature almost uniformly for each active site at the free surface. In contrast, for multi-component nanocrystals, the non-linearity of the Boltzmann factor alters the MC acceptance probability of the active sites according to chemical composition. We observe a slight change of the shape transformation pathway as a function of the energy gap (Figure 2). For increasing temperature, the dissolution rate of atoms in the surface shell is faster and decreases less when the core region is exposed. The surface dissolution becomes more uniform and the effect of etching on the shape of the multi-component microstructure less evident. High temperature (above 300 K) decreases the reactivity difference to the oxidative environment for the various chemical species. On the other hand, low temperature enhances the effect of the energy gap and favors the formation of intermediate shapes with more pronounced concavities.

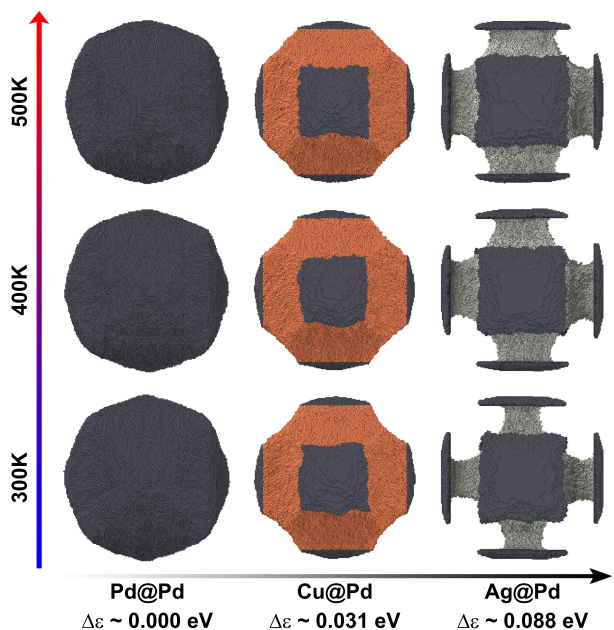

**Figure 2. Variation of the intermediate as a function of temperature and element composition for cube@cube nanocrystals at chemical potential $\mu = -2.20$ eV.** Intermediates of the same chemical compositions are sampled at a constant number of remaining Pd atoms belonging to the surface shell. As temperature increases (row sequences, left to right), and the element energy gap decreases (column sequences, top to bottom) the concavities become progressively less pronounced. Independent of temperature, single-component (Pd@Pd or pure Pd) nanocrystals transform towards a tetrahexahedron polyhedron bounded by {310} facets.

**Chemical sensitivity.** The strength of the oxidative environment affects the dissolution rate of nanocrystal regions with different chemical composition, in a distinctive manner. Function of the bond energy $\varepsilon$, the surface recedes smoothly for a defined range of chemical potential, $-6.0 \leq \mu/\varepsilon \leq -8.0$, whereas, it recedes fast and irrespective of the chemistry of the local environment for larger values.[52] Because the etching reaction proceeds along well-defined crystallographic directions, the progression of the etching process does not change as long as the geometric properties of the particles are kept constant. Importantly, the relative depth of the receding front determines the shape of the intermediate states. The same oxidative environment can result in a range of different surface facets depending on the bond energies of the core and the surface shell. The curvature of the surface envelope changes gradually with the applied chemical potential (Figure 3). The intermediate shape of Cu@Pd rhombic-dodecahedron@cube nanocrystals (Movie M_CuPd_rc) transitions from convex for weakly oxidative environments ($\mu \sim -2.00$) to concave for strongly oxidative environments ($\mu \sim -2.80$ eV). Ag@Pd nanocrystals with larger energy gap show an even stronger progression, including the exposure of the interface between the two crystalline domains for a sufficiently strong oxidative environment (Movie M_AgPd_rc).

Chemical potential and bond energy tailor the formation of new facets by modifying the sensitivity of the different nanocrystal regions to the oxidative environment. The two contributions to the dissolution rate are ideally interchangeable. The same intermediate shapes are achieved either for nanocrystals with small energy gap and high chemical potential (Cu@Pd and $\mu = -2.65\pm0.15$ eV) or for nanocrystals with large energy gap and low chemical potential (Ag@Pd and $\mu = -2.05\pm0.05$ eV). However, the interval width of chemical potential required for the same progression narrows with increasing energy gap. Hence, fluctuations of the oxidative strength (*i.e.*, concentration of the ions) in the liquid bath becomes more significant. Deviations from the expected isotropic behavior can, therefore, occur because of the inhomogeneous environment.

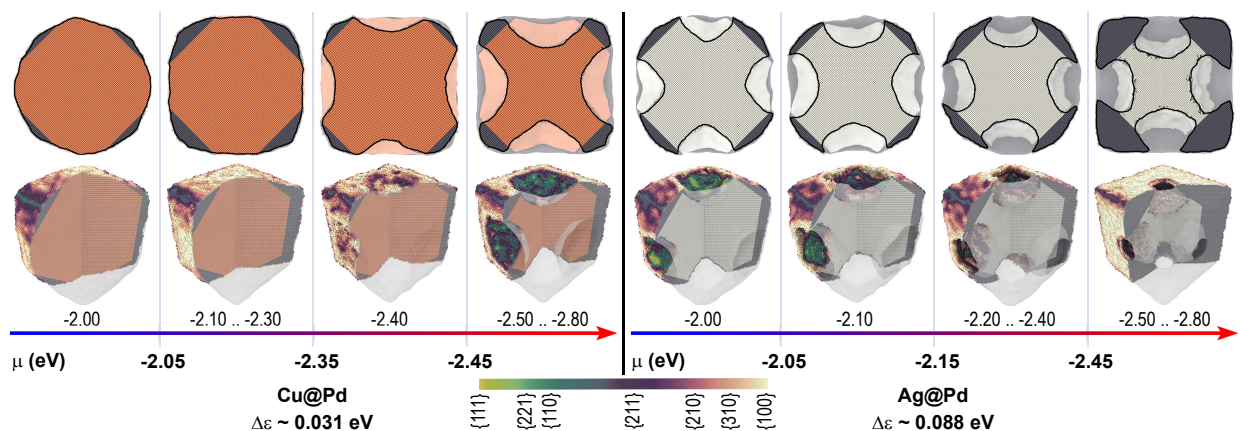

**Figure 3. Continuous variation of the intermediate as a function of chemical potential and element composition for rhombic-dodecahedron@cube nanocrystals.** Cut-section through the nanocrystal center along the [100] zone axes (first row), and three-dimensional view of the nanocrystal without the font quadrant along the [111] zone axes (second row). Atoms are colored according to either the direction of the local surface facet, or the element composition. As the chemical potential µ increases, the exposed surface of the core region progressively varies from convex to concave (Cu@Pd sequence). With increasing energy gap $\Delta\varepsilon$, the inner core@shell interface eventually becomes exposed, facets with high miller indexes are formed and a wider distribution of facet orientations evolve at the surface of the nanocrystal (Ag@Pd sequence).

**Element-specific etching energy and dissolution regime.** For multi-component nanocrystals, the chemical potential imposed by the environment solution sets different dissolution regimes depending on the element species (Figure 4A). In this section, we consider this effect by extending the model. We define the etching energy of an element species as the energy for the diffusion as well the deposition of an atom from or to an effectively defect-free bulk system (Figure 4B),

$$\tilde{H} = \Delta H_{bulk}(C \rightarrow C'). \tag{6}$$

The transformation of the core@shell nanocrystals is thus a function of the etching energies of the element species $\alpha$ and $\beta$ belonging to the surface-shell and core regions, respectively. The energy gap characterizes the different progression of the chemical reaction between the two particle regions. Concavities form on the intermediate surface with a depression depth proportional to the etching energy gap, $\Delta \tilde{H}_{\alpha\beta} > 0.0$. Indeed, as long as the same chemical potential applies to the various element species, the intermediate shapes of Ag@Pd nanocrystals are more open than for Cu@Pd (*i.e.*, $\Delta \tilde{H}_{PdCu} < \Delta \tilde{H}_{PdAg}$, ~1.06 < ~0.37 eV), where

$$\Delta \tilde{H}_{\alpha\beta} = 12(\Delta \varepsilon)_{\alpha\beta} = \Delta E^0_{\alpha\beta}. \tag{7}$$

We now consider distinct chemical potentials for each element species to take into account their specific reactivity to the oxidative environment. The externally applied chemical potential $\mu_e$ (*i.e.*, applied electrostatic potential) is corrected by the standard electrode potential $\mu_0$ as

$$\mu = \mu_e + \mu_0. \tag{8}$$

Depending on the core@shell system, the importance of this correction in predicting the correct shape evolution varies. Transformation of Ag@Pd nanocrystals is insensitive to the correction (Movie M_SEP_AgPd). The standard electrode potential for Ag (0.80 eV [54]) and Pd (0.92 eV [55]) are similar compared to the sublimation energy gap, $\Delta E^0_{AgPd}$ ~1.06 eV. Independent of employing this correction, the etching energy gap,

$$\Delta \tilde{H}_{\alpha\beta} = \Delta(\mu_0)_{\alpha\beta} + \Delta E^0_{\alpha\beta}, \tag{9}$$

approaches the sublimation energy gap value ~1.1 eV. On the contrary, the correction is significant for Cu@Pd nanocrystals, which transform to intermediate shapes that are similar to those observed for Ag@Pd. A large etching energy gap of 0.77 eV results from the smaller standard electrode potential of Cu, 0.52 eV,[56] compared to that of Ag and Pd. The behavior of Cu@Pd without the correction is identical to Ag@Cu [57] with correction applied (Movie M_SEP_AgCu). This behavior can be rationalized by the fact that the corrected etching energy gap of Ag@Cu, 0.41 eV, is close to the sublimation energy difference of Cu@Pd, 0.37 eV. Simulations for Ag@Pd and Cu@Pd with uniform chemical potential are therefore representative of the pathway evolution of core@shell nanocrystals with either large or small etching energy gap (Movies M_AgPd and M_CuPd). In summary, element-specific energies, as considered by the chemical potential correction, do not lead to new phenomena or new intermediate, which justify the simplification in the previous discussions.

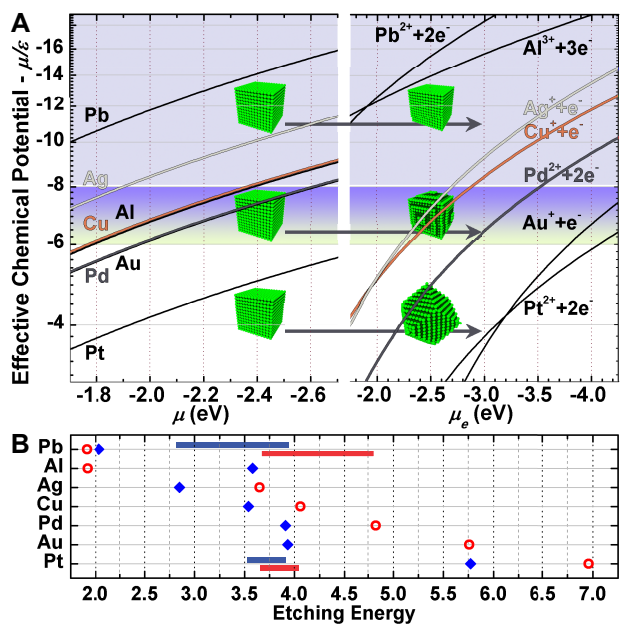

**Figure 4. Behavior of single-component metal nanocrystals as a function of chemical potential $\mu$ and bond energy $\varepsilon$.** A, Cubic crystals transform towards a tetrahexahedron polyhedron exposing {310} facets for $-6.0 \leq \mu/\varepsilon \leq -8.0$. Uniform dissolution (growth) of the surface occurs for stronger (weaker) oxidative environment [52]. The reactivity of the element species (standard electrode potential $\mu_0$) is decoupled from the externally applied chemical potential $\mu_e$ assuming $\mu = \mu_e + \mu_0$ (right plot). B, Etching energy for several element species ignoring (blue full diamonds), and taking into account (red open circles) the standard electrode potential. Ag-Pd and Cu-Pd couples with constant chemical potential $\mu$ (blue bar) have comparable etching energy gap as, respectively, Ag-Pd and Ag-Cu pairs when taking into account the standard electrode potential (red bar).

Further corrections to the chemical potential may be applied to account for additional features of the experimental setup, such as space-time variation of the ion concentration and their diffusion rate. Ligands at the surface of the seed nanocrystals could, for instance, alter the pH of the solution in the proximity of the nanocrystal. Although corrections may be needed to gain a detailed agreement with experimental observations, we believe the proposed model is able to capture the most significant characteristics of the etching behavior as a function of the nanocrystal element and shape compositions.

**Design space for non-equilibrium intermediates.** We explore the evolution of a variety of core@shell nanocrystals systematically to identify and map all accessible intermediates. Figure 5 organizes observed etching evolution pathways as a function of the control parameters core and shell size and shape (cube, rhombic dodecahedron, and sphere), chemical composition, and the chemical potential. We use the etching energy gap as an effective parameter to describe the dependence of the evolution pathway of arbitrary bimetallic core@shell nanocrystals on their chemical composition. Varying these parameters, we observe intermediates that we distinguish as concave polyhedron (CV), pod (PD), strutted-cage (SC), cage (CG) and convex platelet (CX). Nanocrystals transition through multiple intermediate states (intermediates) before they are completely oxidized. Larger differences between shape and etching energies of the core and surface shell regions result in more anisotropic shape transformation. Independent configurations are labeled according to the appearance stage (*I* to *IV*), the core shape of the seed nanocrystal ('c', 'r' and 's' for cube, rhombic dodecahedron and sphere) and the distinct evolution pathways ('a' to 'g'), such that 'IIc/b' stands for the second appearing intermediate shape formed from a nanocrystal with cubic core through the b-th evolution pathway (Figure 5). Concave polyhedrons form by the non-uniform dissolution of the nanocrystal facets progressing from a corner or rather the center to the particle edges (CV – Is/de or Is/fg). At later times, concavities open breaches in the surface-shell region and connect to each other. Throughout the core, pod shapes appear when the concavities are fully exposed to the nanocrystal surface and atoms belonging to the surface are shell restricted to particles' tips only (PD – IIs/d or IIs/f and IIIs/f). Strutted-cage intermediates are obtained when the concavities connect within the core and open to the nanocrystal surface through breaches in the surface shell (SC – IIs/e or IIs/g and IIIs/g). After the complete dissolution of the core region, platelets (CX - IIIc/d and IIIr/d) or cage (CG - IIIs/e or IVs/g) intermediate remain from the surface shell.

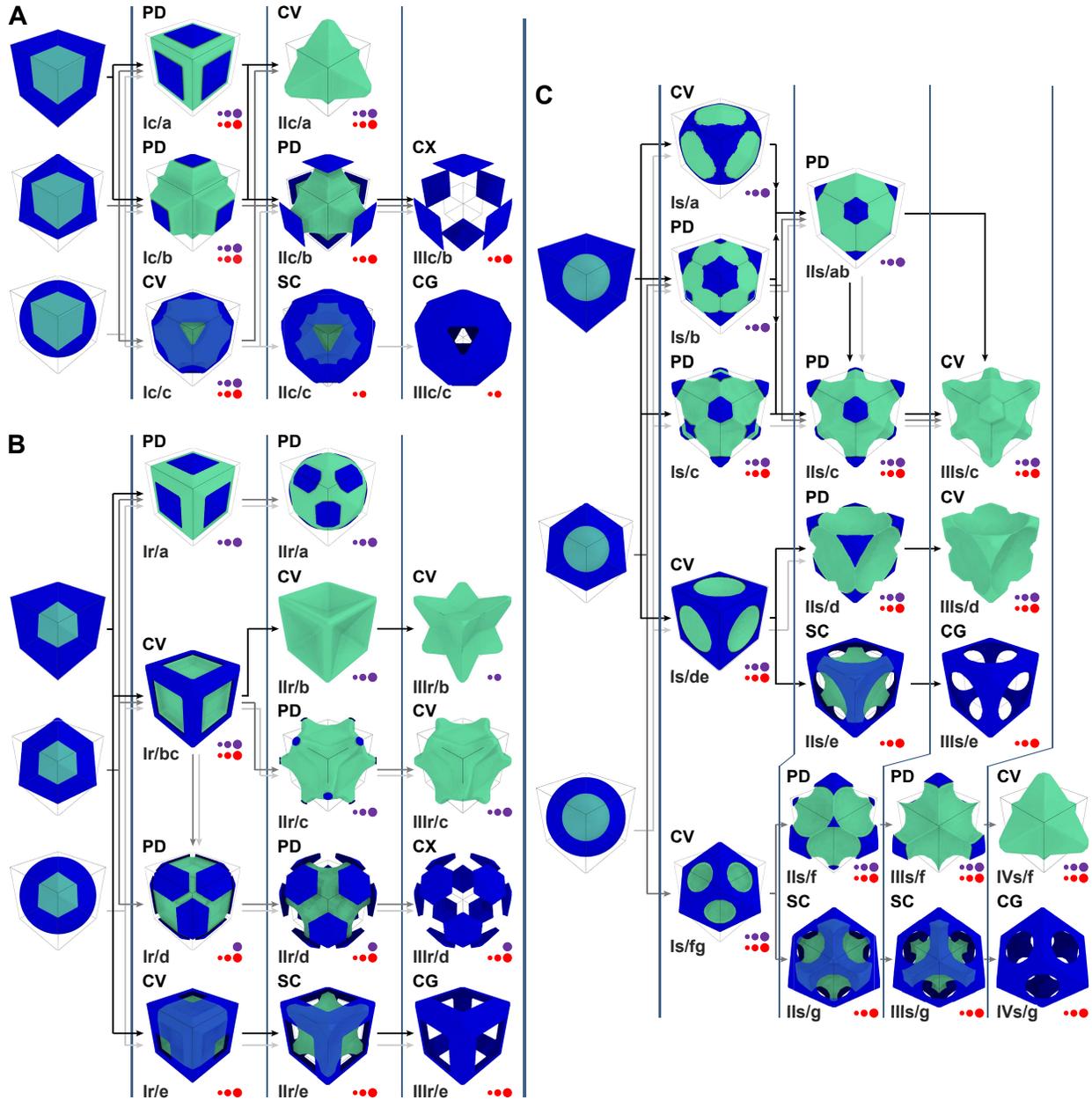

**Figure 5. Summary of evolution pathways of core@shell nanocrystals.** Three-dimensional view along the [111] zone axes of the characteristic intermediate shapes. The chemical elements with higher and lower bond energy than each other are colored the blue and green, respectively. Possible transformation pathways for various temperatures, element composition and chemical potentials are identified as a function of the shape of the core (three groups) and of the surface shell (arrows). Within each group, pathways are listed in order of decreasing chemical potential (*i.e.*, stronger oxidative environment). The intermediate states are marked according to the

core@shell etching energy gap with violet and red circles on the bottom right of each snapshot for small and large energy gap, respectively. Surface shell thickness over particle size ratio of the seed nanocrystals is distinguished by the circles' size from left to right for smaller than, equal to, larger than 1/6. The shapes are distinguished as concave polyhedron (CV), convex platelet (CX), pod (PD), strutted-cage (SC) and cage (CG) at the top left of each snapshot and are uniquely named (bottom left of each snapshot) according to the appearance stage (I to IV, from left to right), the core shape of the seed nanocrystal ('c', 'r' and 's' for cube, rhombic dodecahedron and sphere) and the distinct evolution pathway ('a' to 'g'). See movies for the detailed evolution pathway (Appendix A)

Given that the outer envelope transforms towards the THH for a broad range of conditions,[52,53] the shape and size of the core remain the most important control parameters for fixed chemical composition. We find that the orientation of the legs, struts (or slender fragments in general) and/or surface depressions are a direct consequence of the core shape within the nanocrystal. Distinct shapes are developed by narrowing the surface shell. The thickness of the outer shell controls whether the formation of the THH transition envelope is complete or whether the initial surface shell shape is partially preserved at the time when the core region is first exposed to the surface. {310} facets form completely for shell thickness over particle size (*e.g.*, cube side length) ratio larger than 1/6. Thin surface shells open at an early stage and alter the regression of the outer shape. Either the number of exposed facets or their relative area fraction are then affected, and different transformation pathways for the oxidizing seed nanocrystals with similar configuration are promoted.

An even richer design space is accessible for a pair of element species with a large sensitivity difference to the oxidative environment. A large etching energy gap causes the surface shell to break multiple times over the reaction time and initiate new etching fronts. Nanocrystals proceed to dissolve in a distinct manner at each front. Facets with various orientations are stabilized by the competing regression at the interface between regions of different chemical composition. Extreme conditions (*i.e.*, large etching energy gap and strongly oxidative environment) further open the design space. Cage-like microstructures form from Ag@Pd systems immersed in a strongly oxidative environment. The dissolution fronts never progress outwards from the core region to the surface shell. Hence, the shape of the core region in the seed nanocrystal constrains the orientation of the inner facets that becomes exposed to the oxidative environment (see pairs of shape 'Is/a' vs. 'Ir/a' and 'IIIs/e' vs. 'IIIr/d', among others in Figure 5).

**Thermodynamic stability.** We now investigate whether MD equilibration preserves the layout of the nanocrystal intermediates. Our simulations demonstrate that the initial intermediate microstructure built with a uniform *fcc* lattice structure of lattice constant 0.40 nm is stable throughout thermodynamic equilibration at 300K. Both the stubby and slender regions of the irregular shapes are steady and keep their shape. Legs, struts, and edges comprising the nanocrystal intermediates do not collapse or significantly bend, retaining the crystalline structural continuity. Very few edge dislocations and/or stacking fault defects form. Furthermore, the different atom species do not diffuse over the surface or migrate through crystalline domains. Relaxation of the atom bonds smears asperities at the microstructure surface (Figure 6 top *vs.* bottom row). Weakly bounded atoms stitching out of the microstructure either collapse back to the surface or rarely detach, flying away. Less than 0.01% of atoms per nanocrystal detach from the intermediates due to the stress concentration at the microstructure surface.

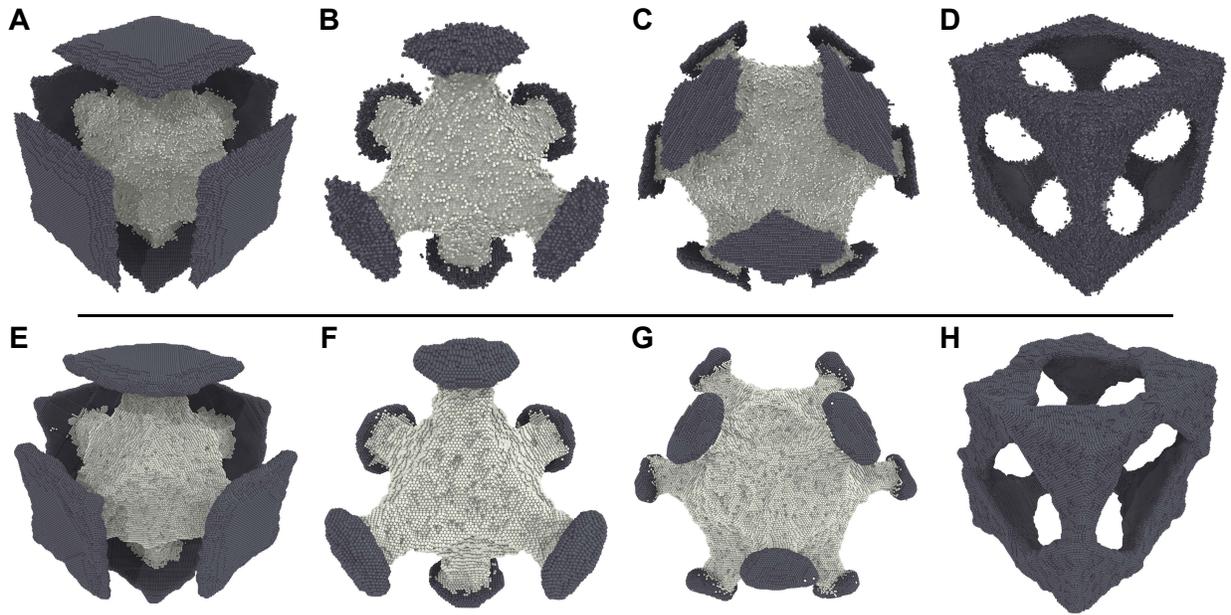

**Figure 6. Thermodynamic stability of intermediate microstructures.** Tension relaxation of the free surface from molecular dynamics (MD) equilibration. Ideally crystallographic as obtained from MC simulation (A to D) and the corresponding MD equilibrated (E to H) intermediate morphologies. Although the characteristic geometric features are preserved, surface atom migration (C vs. G) and residual microstructural deformation (D to H) are eventually observed.

Relaxation of interatomic bonds of under-coordinated atoms affects the lattice structure of small portions of the nanoparticles. Surface tension deforms the particle microstructure according to the structure mechanical anisotropy. Deformations are larger where the slender or sharper portions of the nanocrystals arrange along the stiffer ⟨110⟩ than ⟨100⟩ crystallographic directions. Pd islands at the tips of the ⟨100⟩ particle legs are less sensitive to surface relaxation (Figure 6A and B *vs*. E and F). In contrast, the rhombic regions at the tips of the ⟨110⟩ pod legs tend to compactify by resettling atoms inwards along the ⟨$\bar{1}10$⟩ directions, forming disk-like domains (Figure 6C *vs*. G). Concavities and holes intensify the stress. Cubic cage particles twist by accommodating a few stacking fault defects in the center of the edges (Figure 6D and H). The circular holes left by the etching of sphere@cube nanocrystal at the cubic surface envelope transform towards ⟨110⟩ rhombohedra. The microstructure boundaries shear to equilibrate internal stress to minimize energy.

**Structural and microstructural vibration modes.** Correlations of atom displacements activate microstructural vibration modes. The atom mean square displacement (MSD) calculated from MD trajectories of irregular multi-component microstructures is significantly larger than MSD for compact nanoparticles. In addition to the high-frequency (THz) vibration modes that characterize thermal motion, particle regions with sharp structural rigidity fluctuations trigger long-range microstructural vibrations in the GHz regime. The large fraction of under-coordinated atoms at the surface of nanocrystals increases the atom mobility compared to bulk materials.

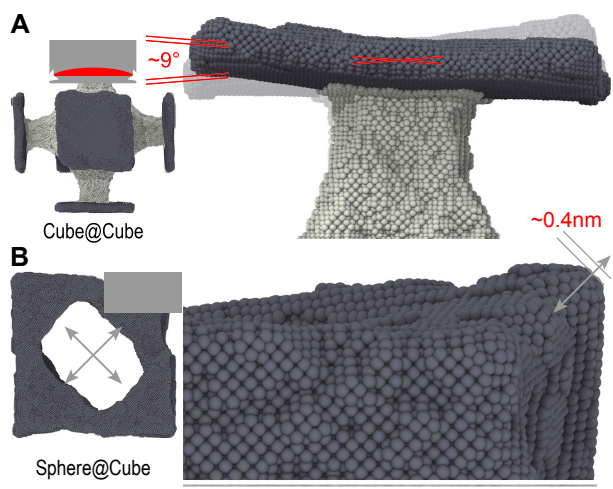

**Figure 7. Microstructural vibration modes.** Thermal vibration of atoms becomes coherent across the nanocrystals. Tails of the pod legs bend (upper line, A), and the cage twist (bottom line, B) at an approximate frequency of 10 GHz (vibration time interval of 0.1 ns).

Convex-shape nanocrystals expand and contract repeatedly over time as the thermal atom vibrations superimpose constructively or destructively. In comparison, distinct portions of irregular concave microstructures vibrate pronounced in space. The wide Pd leg tails of branched nanocrystals oscillate as a rigid body, and the torsion angle of the caged systems fluctuates over time (Figure 7). Both of these low-frequency vibrations raise the apparent MSD by an order of magnitude compared to its reference bulk value. Although the lattice distortion over the nanocrystal is constant over time, it varies locally. Thermal vibration modes are confined in the nano-sized crystals, such that small local atom displacements become coherent throughout the particle. The entire body configuration is, therefore, affected by periodic, collective deformations. Slender particle regions magnify this effect because of the missing body constraint. Indeed, the furthest corners of open shapes show strong displacements not accessible for compact nanocrystals. The absolute atom displacements ($\geq 0.400$ nm) are easily larger than characteristic interatomic distances (~0.389 nm) regardless of the presence of any structural defect.

**CONCLUSIONS**

Chemical selectivity and performance can be targeted by the design of advanced functional nanostructured materials. Applications require controlling geometric features and element composition, simultaneously. In this work, we proposed and explored etching in an oxidative environment of core@shell nanoparticles as an effective synthesis method towards multi-component nanomaterials with non-convex and porous surface features. We demonstrated that a wide set of morphologies can be realized by appropriately tuning the initial seed microstructures (shape, size and element composition) together with the environmental conditions (chemical potential and temperature). Unusual and non-equilibrium surface facets resulted from co-dissolution of crystalline sub-domains, highlighting the potential of this methodology. Non-equilibrium nanocrystal transformation pathways were rationalized and relevant mechanisms affecting the shape evolution discussed. Although surface relaxation in many situations is known to affect thermodynamic stability, our work demonstrates that the concave nanocrystals obtained by the etching procedure are unusually robust. This systematic study provides a reference for future experimental etching studies of multi-component nanocrystals.

**METHODS**

**Numerical simulation of the chemical etching process.** We model chemical etching of metal nanocrystals by means of Monte Carlo (MC) simulations of a lattice gas model.[52,58] Agreement with liquid cell transmission electron microscopy observations for single-component nanocrystals [52] supports the reliability of this approach. In this model, the state of the system, $(\rho_i)_{i \in \Omega}$, is an assignment of atomic occupancies $\rho_i$ to the sites of a lattice $\Omega$. Each site describes an atom of the nanocrystal such that $\rho_i = 1$ specifies a site occupied by an atom and $\rho_i = 0$ marks an empty site. The energy of state $C$ is

$$H(C) = \sum_{i \in \Omega} \left( \rho_i \mu_i + \frac{1}{2} \sum_{j \in \Pi_i} \rho_i \rho_j \varepsilon_{ij} \right), \tag{1}$$

where $\Pi_i \subset \Omega$ is the set of first neighbors of the site $i$, $\varepsilon_{ij}$ is the bond energy for the atom pair $i-j$ and $\mu_i$ is the chemical potential of the atom at the site $i$. The bond energy between atoms of the same chemical species is the bulk sublimation energy $E^0$ divided by the coordination number in a defect-free crystal structure (12 for *fcc* metals).[59] We apply geometric averaging to obtain the bond energy between atoms of different chemical species. The chemical potential $\mu_i$ is a proxy for the sensitivity to the oxidative environment, *i.e.*, the relative concentration of the etchant ions and the applied electrode potential. Unless specified otherwise, we set the chemical potential for each site to the same value.

Etching progresses by MC moves that switch the occupancy of a randomly chosen sites. A MC move from state $C$ to state $C'$ is accepted according to the Glauber acceptance probability [60]

$$k(C \rightarrow C') = w_0 / (1 + e^{-\beta \Delta H}), \tag{2}$$

where $\beta = (k_B T)^{-1}$ with Boltzmann constant $k_B$ and temperature $T$, and $\Delta H$ is the energy gap between $C$ and $C'$. The time-scale factor $w_0$ maps the simulation on a full kinematic scheme such that $k(C \to C')$ is the chemical reaction rate.

Only atoms exposed to the oxidative environment can be dissolved. We restrict, therefore, MC moves to the set of active sites $\Gamma \subseteq \Omega$, where an active site $i \in \Gamma$ is defined as a site that has at least one first neighbor $j \in \Pi_i$ with $\rho_i + \rho_j = 1$. In this sampling scheme, insertion moves $\rho_i = 0 \mapsto 1$ are performed for empty active sites $i \in \Gamma_{\rho_i=0}(C)$ and deletion moves $\rho_i = 1 \mapsto 0$ are performed for occupied active sites $i \in \Gamma_{\rho_i=1}(C)$. To recover detailed balance, the acceptance probabilities for insertion and deletion are modified to [52]

$$k_{\text{insertion}}(C \to C') = \frac{w_0}{1 + \left[ \left| \Gamma_{\rho_i=1}(C') \right| / \left| \Gamma_{\rho_i=0}(C) \right| \right] e^{-\beta \Delta H}}, \qquad (3)$$

$$k_{\text{deletion}}(C \to C') = \frac{w_0}{1 + \left[ \left| \Gamma_{\rho_i=0}(C') \right| / \left| \Gamma_{\rho_i=1}(C) \right| \right] e^{-\beta \Delta H}}, \qquad (4)$$

where the energy difference

$$\Delta H = H(C') - H(C) = \pm \left( \mu_i + \sum_{j \in \Pi_i} \rho_i \rho_j \varepsilon_{ij} \right), \qquad (5)$$

has a positive sign for an insertion move and a negative sign for a deletion move.

**Characterization of single-particle transformation.** Experimental data is necessary to map the number of MC moves on a physical reaction time. We define, therefore, a relative time that increases linearly with the number of attempted moves and reaches 1 once the nanocrystal is completely dissolved. We map the transformation of the nanocrystal over the relative time, and record properties such as the dissolution rate and the shape of the intermediate states. The dissolution rate is defined as the variation of the nanocrystal size per unit of relative time, where the nanocrystal size is computed as the edge length of a cube with a volume equal to the number of occupied sites times the bulk atomic volume. We use compactness [61] as a measure for the nanocrystal shape, which takes values greater than 0 and equals to 1 for a sphere.

**Thermodynamic equilibration of intermediates.** We apply classical molecular dynamics (MD) simulations with the embedded atom method [62–64] implemented in the software package LAMMPS [65] to test for thermodynamic stability of the intermediates. The simulation is performed at room temperature (300K) for a long enough time to ensure equilibration, and then configurations are sampled at constant time intervals. We characterize the lattice structure distortion for each configuration and the corresponding average over time.[66] The structure coherence is evaluated based on common-neighbor [67] and bond angle [68] analysis, whereas the structure deformation is measured by crystal cell and Voronoi cell deformation [69] methods.

**Acknowledgments**

The authors wish to thank X. Ye for discussions and help suggestions. We acknowledge funding from Deutsche Forschungsgemeinschaft through the Cluster of Excellence Engineering of Advanced Materials (EXC 315/2) and support by the Central Institute for Scientific Computing (ZISC) and the Interdisciplinary Center for Functional Particle Systems (FPS) at Friedrich-Alexander University Erlangen-Nürnberg. Computational resources and support provided by the Erlangen Regional Computing Center (RRZE) are gratefully acknowledged.

**Supporting Information Available**

Monte Carlo simulations of core@shell single-crystal nanoparticles oxidatively etched in a redox environment recorded as movies are available as Supporting Information. This material is available free of charge via the Internet at http://pubs.acs.org.

**AUTHOR INFORMATION**

**Corresponding Authors**

*Alberto Leonardi, alberto.leonardi@fau.de
*Michael Engel, michael.engel@fau.de

**Author Contributions**

The manuscript was written through contributions of all authors. A.L. and M.E. discussed the results. A.L. developed computational methods and performed simulations. M.E. supervised the project. All authors have given approval to the final version of the manuscript.

**TABLE OF CONTENTS**

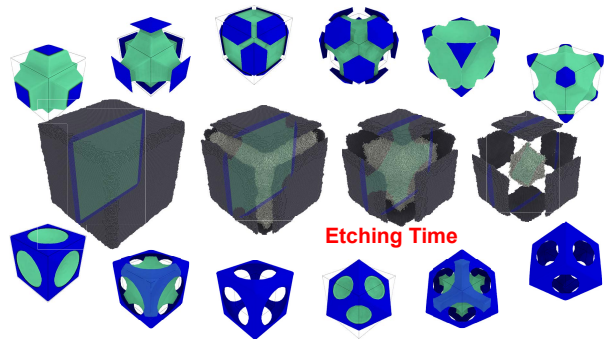